\documentclass[final,5p,twocolumn,sort&compress]{elsarticle}

\usepackage{graphicx}
\usepackage{amssymb}
\usepackage{url}

\begin{document}

\begin{frontmatter}
\title{LCODE: a parallel quasistatic code for computationally heavy
	problems of plasma wakefield acceleration}
\author[binp,nsu]{A. P. Sosedkin}
\author[binp,nsu]{K. V. Lotov}
\address[binp]{Budker Institute of Nuclear Physics SB RAS, 630090, Novosibirsk, Russia}
\address[nsu]{Novosibirsk State University, 630090, Novosibirsk, Russia}

\begin{abstract}
LCODE is a freely-distributed quasistatic 2D3V code
for simulating plasma wakefield acceleration,
mainly specialized at resource-efficient studies of
long-term propagation of ultrarelativistic particle beams in plasmas.
The beam is modeled with fully relativistic macro-particles
in a simulation window copropagating with the light velocity;
the plasma can be simulated with either kinetic or fluid model.
Several techniques are used
to obtain exceptional numerical stability and precision
while maintaining high resource efficiency,
enabling LCODE to simulate
the evolution of long particle beams over long propagation distances
even on a laptop.
A recent upgrade enabled LCODE to perform the calculations in parallel.
A pipeline of several LCODE processes
communicating via MPI (Message-Passing Interface)
is capable of executing multiple consecutive time steps of the simulation
in a single pass.
This approach can speed up the calculations by hundreds of times.
\end{abstract}

\end{frontmatter}

\section{Introduction}
Plasma wakefield acceleration with a particle driver (PWFA)
is a novel method of achieving
high gradient acceleration of charged particles.
The idea boils down to using plasma as a medium to
transfer the energy of the driver --- a high-energy charged particle beam ---
to the witness --- charged particles that follow the driver.
Though the concept dates back to 1985~\cite{chen1985acceleration}
and its feasibility has been well-proven since that time,
experimentally PWFA is still in its infancy,
treated mostly as a phenomenon in need for a more detailed investigation,
not as a mature method ready
to serve immediate high energy physics' needs.

The current flagship PWFA projects are
the already-milestone-setting FACET at SLAC~\cite{litos2014high} ---
the one with the highest energy electron and positron beams ---
and the soon-to-be-launched AWAKE
at CERN~\cite{assmann2014proton, thisissuegschwendtner, thisissuecaldwell} ---
the first one to use a proton beam as a driver.

Studying PWFA relies heavily on numerical simulations,
and calculating the evolution of a macroscopic amount of particles
with a reasonable amount of computational resources
is a great feat of its own.
This raises the need for specialized programs
geared towards performance of PWFA simulations.

A significant performance gain may be achieved by using
boosted frame for the
calculations~\cite{martins2010exploring, martins2010modeling, vay2011modeling}.
Another computationally-effective approach is quasistatic
approximation~\cite{mora1997kinetic, morshed2010efficient, jain2015plasma,
huang2006quickpic, mehrling2014hipace, lotov1998simulation, lotov2003fine},
which allows decoupling the beam simulation time step
from the plasma simulation time step at the cost of limited applicability.
For example, particle trapping by the plasma wave,
sharp longitudinal plasma density gradients or
betatron radiation of the beams
cannot be self-consistently simulated by quasistatic codes.

\section{LCODE}
LCODE is a freely-distributed code for simulations
of particle beam-driven plasma wakefield
acceleration~\cite{lotov1998simulation, lotov2003fine, lotov2013simulation, lcodesite}.
Its main characteristic features are:
\begin{itemize}
	\item 2D3V plane or axisymmetric geometry,
	\item co-moving simulation window that moves with the light velocity,
	\item quasistatic approximation,
	\item fully kinetic, fully relativistic beam model,
	\item optional fluid plasma solver,
	\item field-based kinetic plasma solver,
	\item suppression of small-scale plasma density noise,
	\item automatic substepping,
	\item extensive in-built diagnostics.
\end{itemize}

Automatic substepping --- decreasing simulation step sizes
to match the target precision ---
is implemented in both beam (for low energy particles)
and plasma solvers (for fine field structure areas).

Plasma response calculations are done in terms of electric and magnetic fields,
rather than potentials,
allowing to simulate
non-trivial configurations, like
transversely inhomogeneous,
non-uniformly heated,
non-neutral plasmas
or plasmas with mobile ions of several different sorts.

This unique set of features
allows to perform 
multidimensional parameter scans,
consisting of hundreds and thousands full-scale simulation
runs~\cite{lotov2011controlled, lotov2014parameter}
in a reasonable amount of time.
For example, a single full-scale AWAKE plasma cell~\cite{lotov2014electron}
simulation
takes less than a hundred CPU hours.

\section{Quasistatic algorithm}
\begin{figure}[t]
	\includegraphics[width=\columnwidth]{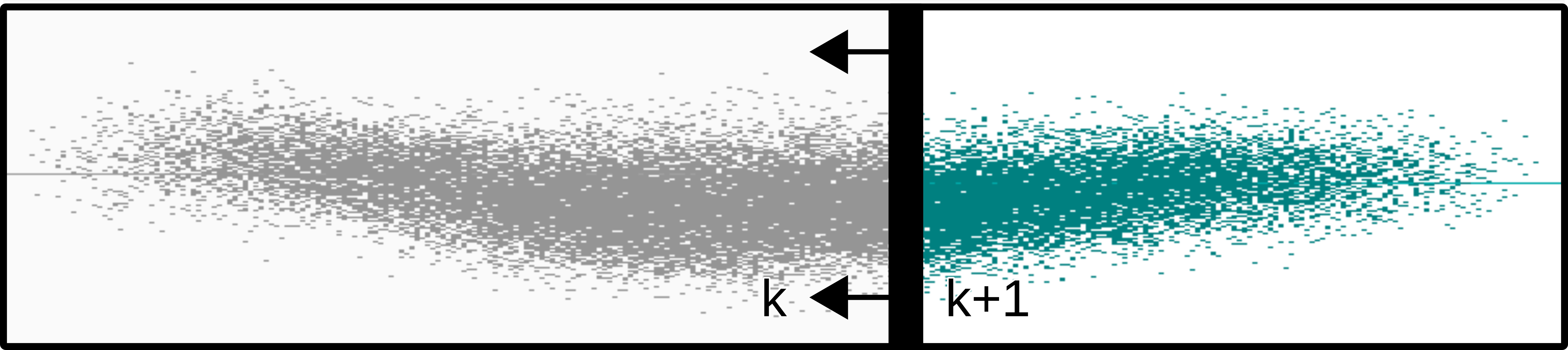}
	\caption{
		Serial implementation of the quasistatic algorithm.
		The beam propagates to the right,
		its next state is calculated from the head to the tail
		in a single pass.
		The picture shows the simulation window
		during the calculation,
		the black bar denotes the current layer position.
	}
	\label{figoneworker}
\end{figure}

Computationally-wise, the particle beam state and evolution
is central to the LCODE operation.
LCODE divides the simulation window into layers, processed from head to tail
(Figure~\ref{figoneworker}).
Ultrarelativistic approximation implies
that in the simulation window, which moves with the speed of light,
the information can propagate only towards the tail of the beam,
so the evolution of each layer depends only on the previous ones.
Internally the beam is stored as a collection of the particles,
ordered by the layer location.
For every layer,
the plasma response to the particle beam
is calculated with iterative algorithms,
then the evolution of the beam is calculated:
each particle's coordinate and momentum get updated
in accordance to the plasma response.
The procedure is repeated for each layer and for each time step.
Slow beam particles can cross the layer borders
and thus participate in several layer calculations during one time step.
From the technical point of view,
at each time step LCODE calculates $(k+1)$th state of the beam
from the $k$th state, layer by layer, head to tail.

\section{Parallel operation concept}
The algorithm described above is, at its heart,
processing a large stream of beam data
and calculating an updated version of the beam in a single pass
along the simulation window.
Thus it is naturally possible to reimplement it
as a stateful stream-oriented transformation.

Here `stateful' refers to the fact that the layer calculation still
depends heavily on the field and particle information carried over
from the previous layers.
Mapping this scheme to, e.g., the actor model
will introduce excessive complexity
because this information would have to be transferred
between different processes frequently.
On the other hand, pinning a dedicated process to a single time step
allows the solvers to simply preserve a part of their state and
carry it over from the previous layer, thus
avoiding interprocess transfer costs.
The only notable downside of such pinning
is the need to adhere to a comparably strict data flow topology.
For example,
as each process possesses unique information about its time step,
the number of processes cannot be easily reduced mid-calculation.

\begin{figure}[t]
	\includegraphics[width=\columnwidth]{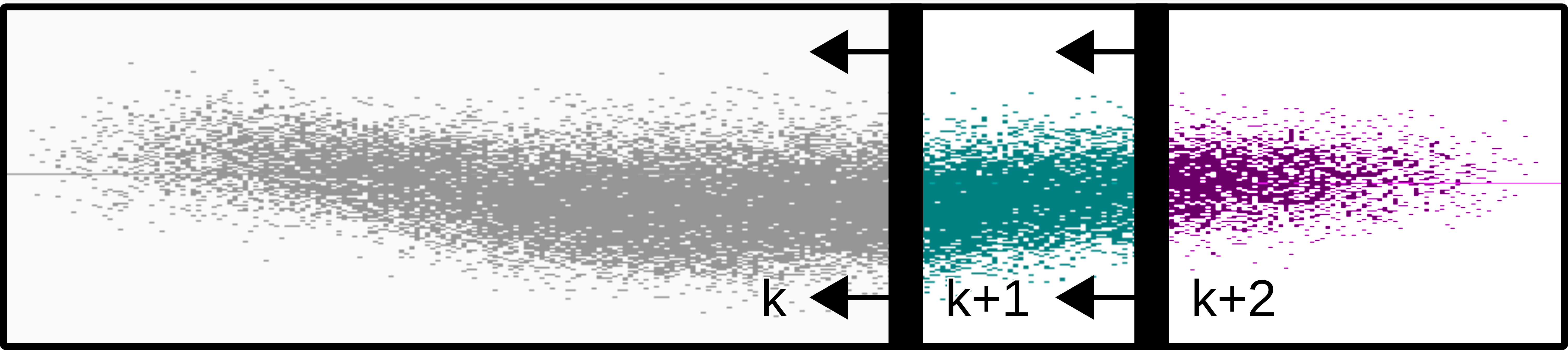}
	\caption{
		Parallel implementation of the quasistatic algorithm.
		The simulation of the $(k+1)$th time step does not have to wait
		until the $k$th step is calculated in the whole simulation window.
		Ultrarelativistic approximation enables a second process
		to start following the first one immediately.
		This way $N$ processes can form a pipeline, calculating $N$ time steps
		of the beam in a single pass.
	}
	\label{figtwoworkers}
\end{figure}

`Stream-oriented' means that
the program consumes and outputs portions of the information
regularly and in small portions:
while it calculates one time step of the evolution of the beam
in a single pass,
it outputs the finished layers of the new generation
immediately after the calculation
and consumes the previous beam generation data as and when necessary.
An output stream of data can be immediately processed
by another instance of this algorithm
as soon as at least one layer
of the preceding time step is completed
(Figure~\ref{figtwoworkers}).
Extending this pipeline to $N$ instances
allows to calculate $(k+N)$th state from the $k$th one
in a single pass.

A parallel computation, designed
as a pipeline of simultaneously executing stream transformations,
has additional benefits in terms of process interoperation.
Most data exchange and synchronization between processes
can be reduced to read/write primitives.
Moreover, a streaming pipeline does not require shared memory,
and is readily scalable to span across multiple hosts.



\section{Parallel upgrade implementation}
A parallel version of LCODE has been developed recently
using the pipeline scheme described above.

This change required to reconsider
how simulations, diagnostics and beam storage
are implemented within LCODE.
Previously the beam information was stored in RAM
and was indeed available in its entirety for the diagnostics subsystem.
In order to facilitate the implementation of the suggested parallel pipeline upgrade
without resorting to shared memory,
LCODE was revised to perform all aspects of the simulation,
diagnostics and beam storage in a single pass.
This `stream-oriented' LCODE is usable as a
building block for the suggested parallel processing pipeline;
the only thing missing is the data flow between the blocks.
The interprocess communication within the pipeline
is implemented using
MPI (Message-Passing Interface).

An interesting caveat worth sharing:
if the beam data consists only of individual particles' information,
the pipeline will stall upon reaching a long region of no beam particles.
A process cannot advance to the next layer
until it is sure that it has received
all of its particles.
In the absence of an `end-of-the-layer' message,
this can only happen when a particle 
from the other side of the region is received.
As a result, all processes have to pause at the edge of the empty region
and traverse it one-by-one.
This effect can be mitigated by exchanging empty layer information
between the processes.

\begin{figure}
	\includegraphics[width=\columnwidth]{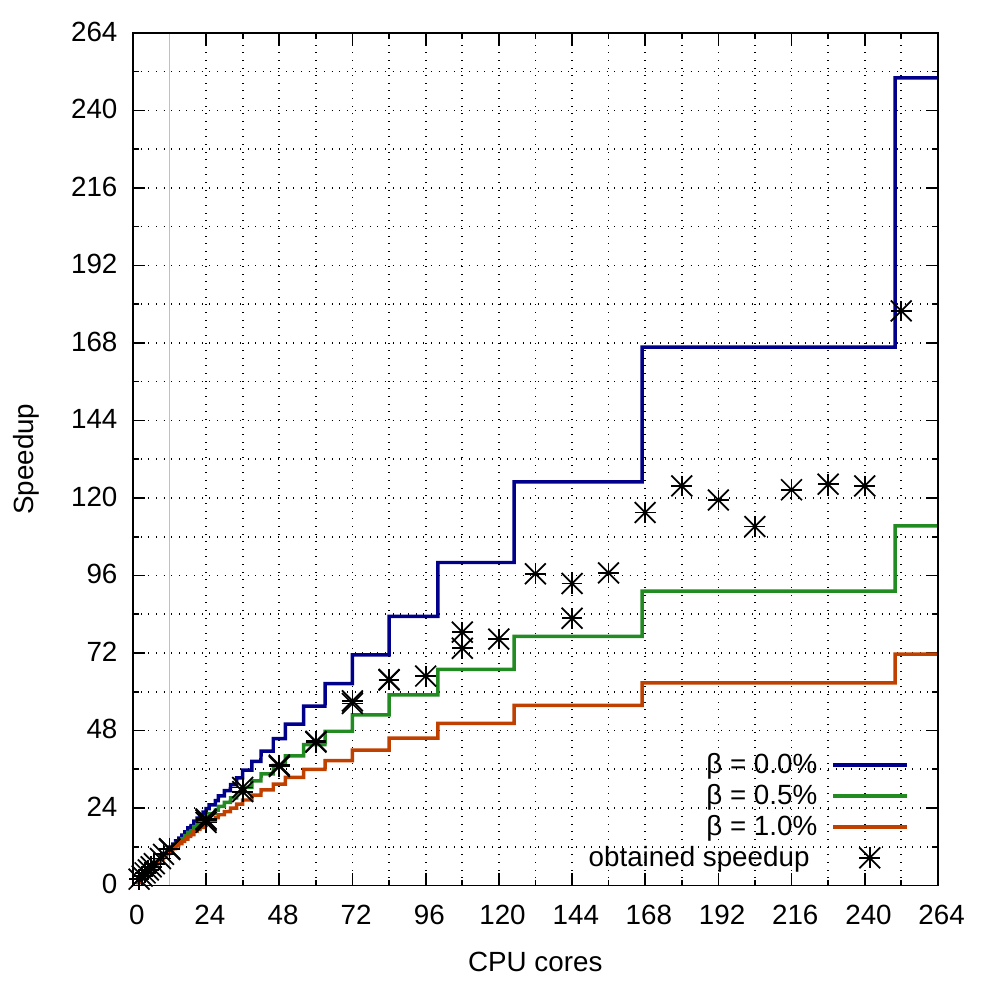}
	\caption{
		Speedup dependence on the number of CPU cores.
		The solid lines show theoretical speedup predictions
		for different values of the
		fraction of the strictly serial execution ($\beta$);
		their step-like shape
		is caused by the fact that the number of time steps is
		not always divisible by the number of CPU cores.
		The results suggest that
		the Karp--Flatt's metric~\cite{karp1990measuring}
		(actually obtained value of $\beta$)
		is less than 0.5\% for most of the runs.
	}
	\label{figspeedeval}
\end{figure}

\section{Speedup evaluation}
The performance of the parallel version was evaluated
on a test task of
simulating the AWAKE experiment
with a doubled plasma cell length (to increase the amount of computations required)
and otherwise reference parameters~\cite{lotov2014electron}.
The obtained execution times were analyzed in accordance with the
Amdahl's model~\cite{amdahl1967validity},
which classifies the calculations as 
either strictly parallel (speeding up by $N$ times with $N$ processes)
or strictly serial (taking the same time to complete for any number of processes).
Results for executions with different numbers of CPU cores
(Figure~\ref{figspeedeval})
suggest a scalability level corresponding to
the fraction of strictly serial execution ($\beta$)
in the order of 1\% or less in Amdahl's model.
To put it simply, using $N$ processes
makes more than 99\% of the calculations $N$ times faster.
The maximum obtained speedup was 177.6~times,
which was obtained by
completing a~5~days~17~hours job in 46~minutes~60~seconds.

\section{Summary}
A parallel version of LCODE was implemented
as a pipeline of processes,
which calculates consecutive time steps of the beam evolution.
This parallel version is freely available on-demand.
Performance evaluation suggests
a nearly-linear speedup for up to
hundreds of processes.
Parallel upgrade allows to obtain simulation results in hours instead of weeks.

\section{Acknowledgements}
The authors would like to thank A.~V.~Petrenko
for his help in preparing this article.

This work is supported by
The Russian Science Foundation (grant~No.~14-12-00043).
Performance evaluation was done at
Novosibirsk State University Supercomputer Center (NUSC).

\bibliography{after}

\end{document}